# Some remarks concerning the Cost/Benefit Analysis applied to LHC at CERN

**Herwig Schopper**

CERN and University of Hamburg

**Abstract**
The cost/benefit analysis originally developed for infrastructures in the economic sector has recently been extended by Florio et al to infrastructures of basic research. As a case study the large accelerator LHC at CERN and its experiments have been selected since as a paradigmatic example of frontier research they offer an excellent case to test the CBA model. It will be shown that in spite of this improved method the LHC poses serious difficulties for such an analysis. Some principle difficulties are due to the special character of scientific projects. Their main result is the production of new basic scientific knowledge whose net social value cannot be easily expressed in monetary terms. Other problems are related to the very strong integration of LHC into the general activities of CERN providing however, interesting observations concerning a new management style for global projects. Finally the mission of CERN (including LHC) is unique since it was founded with two tasks – promote science and bring nations together. No way has yet been developed to assess in economic terms success for the second objective. The main conclusion is that the overall result of the CB analysis, the Net Present Value, although positive for LHC, has a large uncertainty and if used to assess a project needs a detailed discussion. On the other hand partial results can be very useful, for example for the results of education or technology transfer.

## 1. Introduction

During the past decades the necessary funds for research facilities of front research increased considerably. In most cases government funding is required, in particular for basic research. Hence it is not surprising that funding agencies are looking for more ‘objective’ criteria to evaluate new projects, in a time when decision takers hesitate to take the full responsibility and rather try to protect themselves behind a curtain of ‘transparency’. Within such a general tendency cost/benefit analysis (C/BA) has and will become an important instrument since they offer a methodological frame to analyse the value of a project. C/BA has been developed mainly for commercial or industrial projects where the cost and benefits can clearly be expressed in terms of cash. Recently the demand came up to apply C/BA also to scientific infrastructure projects (Martin 1996). In view of the fact that the cost of some projects of scientific facilities are reaching or going beyond the 1000 million EUR this seems to be a justified request.
Scientists should welcome such efforts to accompany the decision making process by a quantitative analysis and thus provide a greater public accountability. However, whereas the cost of a scientific infrastructure can rather objectively be determined, it is a much more difficult task to quantify in accountable terms the benefits of a scientific project and to show that its net contribution to the society is positive. To apply C/BA to infrastructure projects of particle physics is particularly demanding, since the results have a cultural value and the resulting benefits are extremely difficult to quantify.

Indeed some time ago the OECD (1994) studied the socio-economic benefits of Particle Physics and some of the main results of the study were:
-The main reasons for particle physics are cultural; the resulting benefits are somewhat intangible and difficult to quantify.
- Certainly no rigorous cost/benefit analysis is possible. Research into the fundamental workings of the Universe, such as that performed by particle physics, enriches our culture and helps to stimulate other scientific fields.
- Perhaps its largest immediate impact is in the area of education: the questions the field attempts to answer, the sophisticated instrumentation it uses, and the exciting results it provides all act as important stimuli in attracting young people into technically oriented education. A better-educated society is the result.
- but not to neglect secondary benefits in other fields, (technology transfer) e.g. medicine, computing and networking, electronics, low temperatures…
- Better international understanding and setting of models for such co-operations.

Perhaps this view is too pessimistic, but the principle difficulties have not changed. Certainly better methods to evaluate the cost and above all the benefits have to be developed. The most significant progress has recently been made by M.Florio and collaborators (M.Florio, S.Forte and E.Sirtori, 2015). The economic return of a research infrastructure is simply defined as the difference between cost and benefits. Mathematically it is extremely simple, however, the problem lies first of all in the definition of some of the variables and secondly in the way how they can be calculated. The authors are conscious of some of the difficulties and state that this ambitious and difficult task is still going on and the methodology has to be further improved. The Large Hadron Collider LHC at CERN has been chosen as a case study to apply C/BA to a large project of basic infrastructure. The rationale for selecting this project is to investigate a paradigmatic example operating at the frontier of science, for which intangible outputs are particularly relevant.

As an experimental physicist I am not an expert in the field of C/BA methodology and cannot make detailed suggestions in which way to improve the methodology itself. However, based on my managerial experience in creating or operating large scientific infrastructures[1], I shall try to point out some difficulties, perhaps known to a certain extent. This may stimulate discussions and help to improve the method and to recognize its present limitations.

In chapter 2 a short description of the CERN accelerator complex will be given, in chapter 3 some difficulties which are inherent to C/BA applied to basic research infrastructures will be considered and in chapter 4 some problems linked to the special conditions of CERN will be discussed

## 2. CERN and the LHC

The LHC is presently the largest project in basic science and implies an international worldwide cooperation of hitherto unknown extent. However, it is part of a complicated accelerator system and it seems very difficult to separate LHC from the technical, financial and experienced - based environment it is embedded in. To understand this it is necessary to describe shortly how LHC has been created and how it works, indeed how CERN as organisation operates (see also chapt. 4).
CERN has been created originally as a European laboratory after the last World War with the objective to combine the European efforts in particle physics. Since this field of science requires large installations which were beyond the possibilities of individual European states this offered the only possibility to compete with the USA. CERN was considered to be a kind of 'service station' providing for users from the European states:

- facilities like accelerators, computers, machine shops etc with the necessary technical competence
- the overall coordination of international co-operations and management framework
- training for students, scientists and technicians (now also school teachers).

[1] For example Chairman of the Board of Directors of Deutsches Elektronen Synchrotron DESY at Hamburg (1973-1980) , where I was responsible for the proposal, approval and construction of the Synchrotron- Collider PETRA and for the installation of synchrotron radiation laboratories; Director General of CERN, Geneva (1981-1988) responsible for the proposal, approval and construction of LEP, the largest scientific facility of the time; co-founder and president of Council of SESAME, international synchrotron radiation laboratory at Amman, Jordan (2000).

With the first storage ring for proton-proton collisions ISR and later the large electron-positron collider LEP (predecessor of LHC in the same tunnel) and now with LHC CERN has realised unique facilities in the world and has become in practice a world laboratory, although formally still a European Organisation.
Presently CERN has about 11000 outside users from the whole world with only about 100 research physicists among the 2300 CERN staff. Hence science is performed mainly by outside groups ('users') whereas the competence of most of CERN staff is in the fields of many technologies, including networking and computing. The many students from outside universities are a unique source for permanent rejuvenation of CERN users and staff.
The CERN accelerator complex has grown during the last 60 years into a very complicated and multi-connected system. Each of the main machines (e.g. PS and SPS), the largest of its time, has become a pre-accelerator for the following project. For the LHC again the previous machines, starting from a linear accelerator, followed by a booster, the PS and the SPS are used to increase the energies of protons in steps before being injected into the two rings of the LHC for collisions (Fig.1). After acceleration to the required energies particles are stored in two magnetic rings where they circulate in opposite directions and are brought to intersect in a few points. To produce a sufficient number of such head-on collisions between minute particles requires the development of sophisticated technologies. The experiments (ATLAS, CMS, LHCb and others) are installed in these interaction regions to observe the collisions.[2]
The first accelerator to be installed at CERN was a proton synchro-cyclotron for the modest energies of 600 MeV which has been closed down after a very fruitful life of producing excellent data thanks to ingenious experiments. It was followed by fixed-target accelerators, the PS (25 GeV protons) and the SPS (300 GeV protons). In parallel the first proton–proton collider, the ISR, was built. Finally the electron-positron collider LEP was constructed in the 1980ies and closed down in 2001 when it was replaced by LHC.

The main contribution of LEP to LHC is the tunnel. Indeed the 27 km - circumference of the LEP tunnel was only chosen in view of a later installation of a proton collider in the same tunnel. For the purpose of LEP a tunnel with a circumference of 20 to 22 km would have been sufficient[3]. The larger circumference, however, had serious negative consequences for the construction of LEP. About 5 km of tunnel were situated in very bad rock under the Jura Mountain with the consequence that water broke into the tunnel during the excavation, delaying LEP by about one year. In the C/BA analysis for LHC these circumstances were neglected and partly considered as 'sunken cost'. However, without an existing tunnel one could argue that the chances for an approval of LHC would have been marginal, since the geological risks would not have been known. But how to take this into account in a C/BA seems extraordinarily difficult.

[2] Since particles and antiparticles have opposite electrical charges they can circulate in opposite directions in the same magnetic ring. This possibility was used in LEP where electrons and positrons were collided. For proton-proton collisions two rings are necessary.

[3] Originally it was planned to have an electron-positron ring and a proton collider at the same time in the tunnel which would have enabled also electron-proton collisions. However, it turned later out that the space in the tunnel was not sufficient for two machines.

Not only the tunnel but many other parts of the infrastructure of LEP were used for LHC, e.g. existing cryoplants, and assembly halls. The existing machines used as pre-injectors were certainly also a positive condition for the approval of LHC. The LHC and its experiments are benefitting enormously also of the existing infrastructure for computing and networking. Of course, all the additional expenditures for LHC have been determined accurately, but they can hardly be considered to be the true cost of the whole project. Can all this be considered as 'sunken cost'? There is an additional complication. The SPS and also the PS have their own scientific programmes, independent of LHC. The SPS is feeding very interesting fixed target experiments and attached to the PS are several installations for the investigation of antimatter and nuclear physics.
It seems extremely difficult to separate in a project as the LHC with its many cross-connections, the individual items of cost and benefits without taking into account the give and takes of various parts but also the cross-fertilization in experience and know-how.

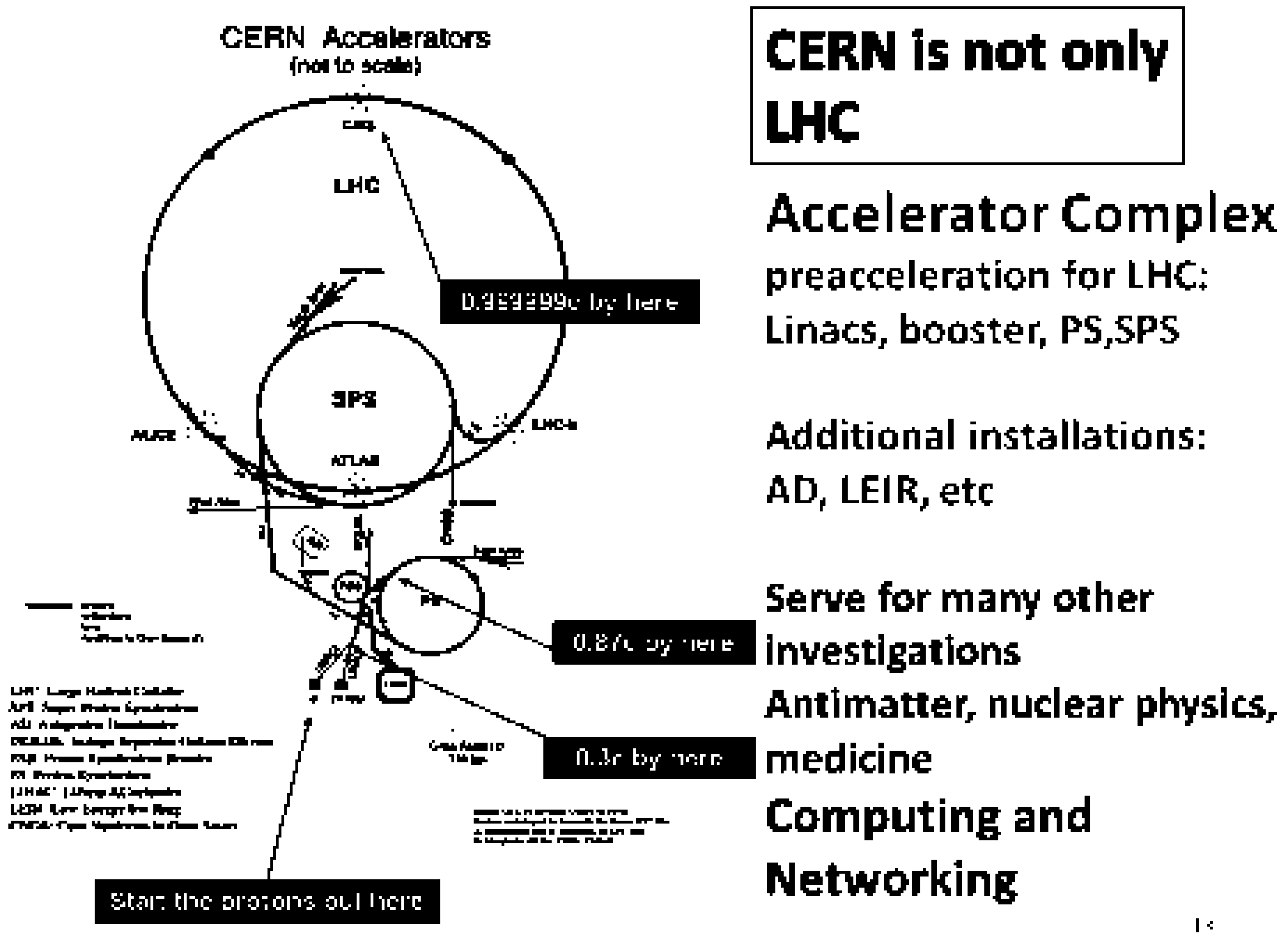

**Figure 1 The CERN Accelerator complex**

## 3. The B/CA method and its problems for basic research infrastructures.

In physics certain rules have been developed as to the choice of the appropriate variables in a minimal problem (Hamilton or Lagrangian method), i.e. variables have to be 'canonical' and in addition certain symmetries have to be obeyed which define the basic properties of a system. From such symmetries certain conservation laws can immediately be deduced. Can similar basic ingredients of the Lagrangian formalism (Dréze and Stern 1987 and 1990) be exploited in C/BA?

I have the impression that the definition of concepts (variables) is still one of the most challenging problems. Indeed Rouse *et al.* (1997) have argued that the main difficulty in applying C/BA in the research domain is not in calculation *per se*, but more in the identification of benefits and costs, which implies a definite definition of variables. They also see a certain difficulty in the attribution of benefits and costs to a project under assessment.

To discuss in some more detail the problems which arise I am following closely the procedure introduced by M. Florio, S. Forte and E. Sirtori (2015 and 2014). The aim of the C/BA is to calculate the *net positive value* NPV of an infrastructure over a certain time horizon. NPV is simply the difference between benefits and costs valued at shadow prices and discounted. If the benefits are higher than the cost and hence NVP is positive the project is estimated to produce a net positive contribution to social welfare. This figure is considered to be the main result of C/BA. The authors take into account five categories of benefits for fundamental research infrastructures (the variable $A$ is more relevant for applied research and is neglected) and they also consider five groups of costs. Thus the NVP is calculated by the formula (meaning of variables see Table 1)

$$NPV = [S + T + H + C + A] - [K + L_s + Lo + O + E] + B_n \quad (1)$$

where the first bracket contains the benefits and the second the cost, both to be discounted for the different times at which they occur. $B_n$ is the sum of two types of '*non-use values*' related to research discoveries $B = QOV_t + EXV_t$ where the quasi-option value $QOV_t$ includes any future but unpredictable economic benefits. It is intrinsically uncertain and therefore not measurable, hence simply assumed to be non-negative and neglected. $EXV_t$ is an *existence value* related to pure new knowledge per se and is proxied by stated or revealed willingness to pay for scientific research. In environmental C/BA (Pearce et al., 2006) it is the benefit of preserving something known to exist. As will be discussed later this treatment of is one of the main uncertainties affecting the overall value of NPV. All variables are treated as stochastic and conditional to joint probability distributions of several parameters.

It cannot be the objective of this article to discuss in detail the method of C/BA nor the evaluation of all the components of the *Net Positive Value*. Florio and collaborators have made a courageous and seminal step to develop the method, however, they state themselves that the task has not yet been fully fulfilled. We shall discuss a few special issues which will indeed demonstrate some of the still inherent difficulties of CBA when applied to basic science infrastructures. The hope is that this discussion might lead to further improvements of the CBA method.

### 3.1 Scientific knowledge output

The knowledge output S (equation 1) is one of the variables which is extremely difficult to quantify, but one of the core benefits of LHC. To evaluate the benefit of scientific output of research even from the scientific point of view alone poses problems. The number of publications is certainly one criterion. If a scientist or laboratory has no publications at all within a reasonable period this should be taken as a very negative sign. However, the number of publications does not say anything as to the quality of the research and also the frequency of citations has to be interpreted with great care. Several bibliometric techniques, all kinds of indices, have been developed and became fashionable. However, without proper interpretation they can be totally misleading[4].

[4] See for example the recent article by Roberto Piazza, EPN 46/1,2015, page 19

Scientometrics methods have been developed for the purpose of C/BA to analyse the knowledge output of research infrastructures (Pinski and Narin 1976, Martin 1996). Its value is “empirically proxied by the sum over time of the present value of papers (articles, preprints, etc.) signed by scientists at the research infrastructure, the value of subsequent flows of papers produced by other scientists that use or elaborate of the research infrastructure scientists’ results, divided by the number of references they contain, and the value of citations each paper receives, as a proxy of the social recognition that the scientific community acknowledges to the paper”. To improve the methodology Florio et al (2015) propose several additions. For example they use a model by Carrazza, Ferrara and Salini (2014) to forecast the trajectory of papers and citations related to the LHC experiments. Also a very thorough analysis concerns the distinction between LHC users, outputs produced by non-LHC scientists citing the work signed by LHC users, outputs produced by other scientists citing the mentioned papers, and so on.
However, in spite of these very interesting efforts the evaluation of the scientific productivity (even without trying to express its value in money) of a person, an organisation or its infrastructure is very uncertain because of a few principle difficulties, especially true for particle physics.

Here I can give only a few examples. A particular problem in experimental elementary particle physics is the large number of authors. It is obvious that not all authors contribute in the same way to a particular publication. Some papers deal with instrumentation, other with the analysis of data. At the time of the LEP experiments I proposed that a collaboration writes a few publications explaining the concept of the experiment and the design with all authors signing these papers, whereas later publications with experimental results should be signed only by those who contributed to the particular analysis. For complicated social reasons this was refused. If all authors sign all publications I proposed that at least the alphabetical order should be changed which has the practical advantage that questions can be directed to the directly concerned scientists. Also this proposal was rejected, but recently in some LHC publications a contact person is given. This whole problem with the large number of authors in international collaborations does not only appear in particle physics but also in other fields where co-operations become more important. Dividing the number of publications by the total number of authors, in all cases, seems meaningless.

Using the number of citations in order to evaluate the benefits of a publication poses another difficulty. Sometimes a wrong experimental result is quoted extremely often since it may be in contradiction with all accepted knowledge. Once I was informed about a comparison between different laboratories based on publications and citations. One laboratory did particularly well, but the evaluators had not noted that the extraordinarily high number of citations was due to a wrong experiment whose results contradicted all previous experiments and eventually turned out to be wrong.
The number of citations may also depend on the ‘fashions’ which prevail at a certain time. Let me give just two examples. The Nobel Prize winner Peter Higgs mentioned once in a discussion at CERN that according to scientometrics he would have difficulties to obtain today a professorship. His article proposing a special mechanism to break the gauge symmetry in the standard model of particle physics was published several decades back, was practically not quoted for many years, but eventually Higgs received the Nobel Prize together with F. Engler in 2014 after the Higgs particle was discovered at the LHC. A similar story is told by S.W. Hell (Nobel Prize for Chemistry 2015) whose unconventional ideas seemed to be completely against all knowledge at the time when he made proposals for better microscopy.

In conclusion even in the purely scientific domain the number of publications and their citations are disputed as a means of evaluating scientific benefits and one talks even about the 'high-impact-factor syndrome' (see A.Pawlak, 2015 and C.M.Caves 2014).
Even the number of Nobel Prizes is not a satisfactory tool to evaluate a project or laboratory. One reason is the rule to award it for each field to a maximum number of three scientists in one year. Since in many scientific domains experiments are being done by large collaborations these are excluded from the Prize. The rule favours theoretical work which can still be done by individuals or in small groups[5]. Thus the Nobel Prize for the Higgs discovery was awarded to the theoreticians and not to the two experimental collaborations which detected it.

An even more serious problem concerns the translation of the scientific results of the publications into an economic value. Florio et al (2015) use the marginal production cost concept: the value of each knowledge output is proxied through its opportunity cost, i.e. the value of time devoted to produce the output. Under the assumption that the opportunity cost of knowledge outputs produced by the LHC researchers is proxied by their average hourly compensation, the economic value of papers is equal to their production costs, which is the cost of scientific personnel employed at the LHC and its experiments. In a similar way the value of citations is estimated in terms of opportunity costs taking into account the time for reading and understanding them. If it is conservatively assumed that on average one hour is needed to read and cite a paper, the average hourly salary is taken as an estimate of the social value of one citation.
To use 'opportunity costs' to evaluate a scientific achievement seems doubtful. In real life no correlation between the value of a scientific result and the time spent to obtain it can be established. The social value of a publication does certainly not depend on the time spent, nor on the salaries of involved scientists. How can a single discovery like the Higgs boson be evaluated by calculating the time and multiplying it with the salaries of the involved scientists?

Florio et al are well aware of such problems and therefore they treat all the variables entering in the evaluation as stochastic variables. The resulting mean benefit of knowledge output is calculated as S = 280 million Euro (equation 1). This figure is negligible compared to other benefits evaluated by C/BA (see Table.1) which determine alone the final result of C/BA. The increase of scientific knowledge is, however, the main objective of a project like LHC. This indicates that the C/BA method cannot really evaluate the benefit of new scientific knowledge, as all other methods fail to do it in a quantitative way.

### 3.2 Technological spin-off

Discoveries in basic science have very often had tremendous consequences for new technologies and thus have changed society. For example the discovery in the 19th century that electric and magnetic phenomena are only two different expressions of the same fundamental force, the electromagnetic force, were the basis for the whole electrification including electric power, radio, television and modern computing. It took however, more than 100 years for this development to bear fruit. Whether the recent discovery at CERN that this electromagnetic force and the weak nuclear force are only two components of a more fundamental force, might have similar consequences for technology remains to be seen by future generations. Is there a reliable method

[5] Even for theoretical physics one can argue that in many cases the success is due to the contributions of many scientists (see Schweber S.S., (2015), European Physical Journ.,40, 53)

in C/BA to evaluate such benefits of particle physics which usually become visible only after long times?

It is easier to evaluate technological spin-offs by contracts which have more or less immediate positive effects. Indeed this kind of technology transfer may be, after the scientific benefit, another important social gain of a project like the LHC. As explained above (chapter 2) the technological competence of CERN is crucial for its scientific success. The realisation of an infrastructure like LHC and its experiments has led to many new technological developments giving rise to considerable technological spin-off. Only a very small part of it has been protected by patents, trademarks or copyrights.
Indeed an essential part of knowhow transfer is achieved by contracts to industry, in particular when it comes to cutting-edge technological projects. This is due to a special but typical procedure, which CERN applies in collaboration with industry. If industry is asked to provide a product which requires a considerable effort of technical development, than CERN, in most cases, builds a prototype and only when this corresponds to the expectations of CERN and after it has been proven that the production is possible, a contract will be signed with a firm. After the contract has been adjudicated all the knowhow which has been accumulated during the development of the prototype will be made available to the firm. During the execution of the contract CERN experts will continuously stay in contact with the firm and intervene in case of problems. This procedure contrasts considerably with the method often practiced by other organisations or government agencies which after adjudicating a contract simply wait for the final product hoping that it will correspond to the expectations and be delivered in time. The CERN procedure is essential to avoid large cost overruns and delivery delays.

This kind of technology transfer by contracts can be relatively easily evaluated and several studies have been made at CERN in the past. Schmied (1975) (covering essentially the construction period of SPS) and Bianchi-Streit et al. (1984) (construction of LEP) interviewed firms and asked them how much their turnover increased and/or how much production cost saving they achieved thanks to the experienced gained by the CERN contract. They defined an economic utility/sales ratio index which is the increase of the turnover (or of cost savings) divided by the value of the CERN contract. Schmied found an ‘economic utility/sales’ ratio in the range of 1.4 (for precision mechanics) and 4.2 (for electronics, optics, computers) with an average of 3, whereas Bianchi-Streit et al. calculated values ranging from 1.7 (cryogenics and superconductivity) to 31.6 (precision mechanics), with an average of 4.2.
These methods have been extended and refined by Florio et al (2015) in an impressive way. By taking into account the benefits due to the supply chain, computing software and other transferred technologies they arrive at a mean value for the technical spillover T= 5400 million Euro (equ.1). However, they remark that a more optimistic scenario would have reached 8.6 billion Euro, still being conservative as far as unknown benefits are concerned.

Such studies may be helpful to provide an indication as to the value of contracts adjudicated by CERN to firms. However, does such a ‘utility factor’ really represent the overall benefits obtained through a contract? The utility factor is based on the immediate profit, but other knowhow may in the long run be even more beneficial to a firm. Several examples could be given for how the basic knowhow on fabrication procedures or just learning how to keep tolerances completely changed the kind of products of a firm. For example a firm constructing iron sheet furniture became a major producer of electromagnetic waveguides, the main difference of the two products being the tight geometric tolerances. Another firm became a major European supplier for vacuum containers for liquid helium by learning the adequate welding techniques.

Many innovations in electronics and networking, very often achieved by CERN outside groups, have not led to contracts and hence could not be taken into account in these studies. Sometimes there are even curious spin-offs for which it is difficult to decide whether they should be considered as of economic or cultural values. A recent example is the development of a device to recuperate and preserve historic voice recordings. Some physicists participating in ATLAS used some of the precision optical tools used to construct and analyse particle trackers, to measure and image grooves on wax cylinders and discs. Such digital representations made it possible to hear and preserve the voice of Alexander Graham Bell, the inventor of the telephone (A.G.Levine 2015).

It is sometimes argued (Boisot et al.,2011) that the more the research infrastructure deals with fundamental research, the deeper is the uncertainty about which technological innovations will spur from it and to what extent they will spread. Additional benefits, completely unknown and unimaginable today, may still arise in the future. One may try to incorporate such residual benefits in some 'non-use vale of discovery' (see chapter 3.5).

### 3.3 Human Capital Formation

Human capital formation concerns mainly training and education at various levels. These are intimately linked with research. Without research the material for teaching at all kinds of levels would get very soon obsolete. Or should we still teach our pupils that the earth is at the center of the universe? On the other hand without teaching and training research would lack very soon good young people. Although education was not one of the main original tasks of CERN over the years it has become an important and multifaceted part of its programme. Although primarily a research organisation CERN offers an excellent environment for training. Indeed in an OECD study (OECD 2014) it is stated that "the intellectual environment at high-energy physics laboratories is exceptional, and is probably comparable to that of the most innovative high-technology companies".

Only a few of the training activities of CERN can be mentioned. For scientists at the post-doc level a general Academic Training Programme has been installed. In addition several more specialized 'CERN schools' are organized, e.g. School of High Energy Physics, School of Computing and an Accelerator School. A summer student programme attracts each year many under-graduates. A special programme for the education of school teachers has been established which in cooperation with UNESCO could be extended even to non-Member States of CERN, for example to African countries. Whereas most of the schools are taught in English, broken English being the universal language of physicists, the courses for school teachers are partly given in their home language in order to facilitate the transmission of knowledge to the school children. When the German Federal President Gauck visited CERN recently he expressed the opinion that the teaching and training programmes of CERN are an essential part of its social benefits.

How can these educational benefits be evaluated in a CB analysis? Can they be associated to a special project like the LHC? Sometimes they are simply considered just as 'cost' giving no benefit, in other cases they are designated as some kind of internal cost. Completely neglecting them would certainly not be justified.
Florio, M., Forte S. and Sirtori, E. (2015) adopt the following procedure: "…in order to quantitatively estimate the economic effect of human capital increase, in terms of incremental earnings, we need to hark back to the evaluation techniques of educational economics. Thus, we need to derive the marginal effect of human capital formation on the earnings gained by former LHC students and expected to be gained over their entire lifetime. The socio-economic effects of human capital formation produced by a research infrastructure like the LHC can be evaluated by analysing the marginal increase of earnings acquired by former LHC students, as compared to salaries that would have been earned anyway, without the educational experience offered by the LHC". In carrying out this ambitious programme the authors make a very careful study as to the beneficiaries tracing the careers of the students and employees of CERN over many years. The detailed findings are quite interesting and the final result is a mean human capital formation (equ.1) of H = 5.5 million Euro .
Such a result is certainly justifiable from a purely economic point of view being based exclusively on incremental earnings, but other aspects could be considered as equally important as will be discussed in the next sections.

### 3.4 Cultural benefits

The main cultural benefit of particle physics is new knowledge about the structure of matter and the development of the universe at its very early stages. However, in the framework of C/BA this is considered to fall into the class of existence values and will be discussed in the section 3.5. For methodical reasons in C/BA only those benefits to the general public are considered which can be quantitatively evaluated. A new term, the 'cultural capital' is introduced (Throsby, 1999 and 2001) and as the most promising valuation technique which is offered to decision makers and sponsors, is the willingness to pay (WTP) method (Snowball, 2008). For museums this value can be calculated from the number of visitors and the entrance fee, but Florio et al (2015) introduce a much more sophisticated procedure for LHC. About 100 000 visitors come to CERN per year, in addition there are travelling exhibitions visited by 30 0000 to 70 000 people per year. These figures are projected to the year 2025. Since all these visits are free of charge a WTP figure cannot be deduced so easily. A method developed mainly for museums and based on zonal travel cost is applied (Clawson and Knetsch 1966).
An additional contribution to the cultural benefits comes from the use of mass media. From an estimate of the number of TV viewers and the number of newspaper articles reporting about LHC events and users of CERN WEB sites, a quantitative benefit can be estimated by multiplying the respective time spent with the average hourly value of per capita Gross Domestic Product.
All these investigations give for LHC then mean present value of the total cultural benefit to the general public of C = 2100 million Euro (Equ.1). The largest share in this estimate comes from the use of mass media.

Although this result seems quite impressive it has a very limited meaning due to the limitations of C/BA. Some of the most important and relevant consequences which are difficult to quantify are not taken into account. To mention just one example - one of the most significant consequences of LHC is that the corresponding outreach (in particular the publicity around the Higgs discovery) has raised the interest for natural sciences among young people enormously and without doubt will have the consequence that more will go into natural sciences and become engineers, mathematicians, physicists, chemist, biologists, digital experts, etc. In view of the many global problems which can only be approached by scientific methods this is another social benefit which cannot easily be evaluated in financial terms.

### 3.5 Quasi-option value and existence values

The non-use value $B_n$ consists of two parts (Equ.1), first the quasi-option value which relates to the benefits of future discoveries. As it is intrinsically uncertain, hence not measurable, it is simply set to zero. One does not see an easy way how to improve the C/BA method in this respect, but is it justified to completely ignore this benefit which could be of overriding importance?
The usefulness of the World Wide Web was at the time of LEP not foreseeable, whereas it is considered today sometimes as one of the new paradigms of society. The LHCGrid, the much more powerful successor of the WEB, has been developed for the LHC experiments. Whether its relevance for society will be as important as the WEB cannot be predicted, but certainly its quasi-option value will be greater than zero.

The existence value is linked to the value of pure new knowledge without any direct economic benefits. It is estimated by the willingness to pay for scientific research (see chapter 3.4). Florio et al (2015) conducted a complicated survey among the population by asking them how much they would be willing to pay to keep the LHC running independent of what the possible future use could be. The average estimate for the existence value evaluated in this way came out as EXV= 3.2 billion Euro. The authors are aware that this result may be controversial but they find it in the range of other Willing-To-Pay studies for other cultural goods.

### 3.6 Costs

Since all expenses of CERN are well documented one might think that the estimation of the total cost of LHC is relatively easy. However, some difficulties appear, e.g. many contributions from Member States as well as non-Member states were made in kind whose value is sometimes difficult to estimate and the border between the infrastructure of LHC, the LHC experiments and other activities are sometimes not well defined. Florio et al (2015) made an impressive effort to estimate the costs as defined in Equ.1 which required the introduction of some new methods.

**Table 1. Expected Values for Benefits and Cost (probability functions)**

| Name | Typ | Mean value | Percentage of total benefits |
|---|---|---|---|
| Variable Equ (1) | **Benefits** | In $10^9$ EUR | % |
| S | Scientific knowledge | 0.28 | 1.7 |
| T | Technological spill over | 5.4 | 32.8 |
| H | Human capital formation | 5.5 | 33.3 |
| C | Cultural benefits | 2.1 | 12.7 |
| B | Existence value | 3.2 | 19.5 |
| | **Total benefits** | *16.4* | *100* |
| | **Total cost** | *13.5* | |
| NPV | ***Net Positive Value*** | ***2.9*** | Monte Carlo error (3 σ) <2% |

However, one assumption influences the overall picture decidedly, that is how the cost of CERN general administration, general expenses and of administrative and technical personal is attributed to LHC. Only 10% are attributed to LHC. In a counterfactual history without the LHC, it is assumed that these costs would have been borne by CERN anyway. If this assumption would be changed to 75% instead of 10%, the total cost would rise by several billion EUR and change the overall picture considerably.
With the more favourable assumptions of 10% of the common costs charged to LHC the final estimate for the mean total cost of LHC is 13.5 billion EUR as shown in Table 1.

## 4. Discussion of Overall Results of C/BA for LHC

Florio et al (2015) have calculated the *Net Present Value* of LHC by a Monte Carlo simulation for all the 19 stochastic variables used and obtain an expected *Net Present Value* of around NPV = 2.9 billion EUR (Table 1). The probability that the benefits exceed the costs is calculated as 92%. This is certainly a very positive result for the social value of LHC.

However, in spite of the considerable effort which went into the extension of the C/BA method for basic science infrastructures this overall result has to be interpreted with great caution because of the various assumptions which were applied and discussed above. To emphasis this only the most relevant assumptions introduced in the C/BA will be mentioned here again.

As shown in the Table 1 the scientific publications originating from work at LHC contribute less than 2% to the total benefits. This certainly does not do justice to all the results obtained already by now by the LHC experiments and in particular the discovery of the Higgs boson which is recognised as a major success in particle physics and was worth a Nobel Prize. As has been discussed in more detail in chapter 3.1 the evaluation of new basic scientific knowledge on a purely economic basis seems impossible.

The second great uncertainty comes from the intimate integration of LHC in the overall CERN activities. Although very careful studies have been made how to attribute certain costs to the LHC a certain arbitrariness seems unavoidable. The most serious case concerns the general administrative cost as discussed in chapter 3.6 of which the authors are well aware. The resulting uncertainty is about of the same magnitude as the final *Net Positive Value* and demonstrates the intrinsic problems of a C/BA for a laboratory with several closely interwoven activities.

The main conclusion is that the *Net Present Value* of the CBA should not be considered as the only and most important result. The results for individual benefits and costs are, on the other hand, extremely interesting and provide a wealth of information about the execution and benefits of a large project (e.g. education, technology transfer). Needless to say that more research concerning further improvements of the C/BA method would be very useful.

So far only those elements which were taken into account by C/BA are considered. However, in order to evaluate the social value of a project in basic research and being realised by an international organisation like CERN some additional criteria should be considered, although admittedly it is not evident how they could be included in a C/BA.

## 5. Special benefits in the case of CERN

In the preceding chapters some of the difficulties of C/BA applied to basic research infrastructures where considered. Also some problems linked to the fact that LHC is so intimately linked to the overall programme of CERN were touched upon. However, a few items have not been taken into account at all and for good reasons – they are very specific to CERN or one does not know how to integrate them methodically into C/BA. Nevertheless they represent a social values comparable or even higher to other benefits taken into account.

### 5.1 The cultural value of basic research, a main ‘product’ of CERN

The main objective of CERN is to produce knowledge of an unexplored territory, the infinitely small. Surprisingly the exploration of the microcosm turns out to be intimately linked to the exploration of the macrocosm. To answer questions like: what is matter, what are we made of, what are the fundamental forces dominating nature has immediate implications for the understanding of the development of the universe. Today we think that the universe started with a 'big bang', an extremely high concentration of energy. In the CERN colliding rings particle collisions are produced which result in energy concentrations similar to the ones which existed about a billionths of a second after the big bang. Hence we are exploring the properties of matter as it existed at the very beginning of the universe. One day the essence of this new knowledge will be taught at schools to our children or grandchildren. Without this kind of fundamental research one would still teach that the earth has the biblical age of about 4000 years and that matter consists of the 4 elements of Aristoteles.

Such fundamental questions touch even on the relation between science and spirituality. Various visits of the highest representatives of various religions at CERN provided some occasion to discuss questions like: can there be a conflict between science and religion. Indeed when Pope John II visited CERN we agreed there cannot be such a conflict. The main reason is that different kinds of truth are valid. In science the ultimate test are experiments which can be repeated at any location at any time. Whereas in religions the origin of truth are revelations. Science and religion and also ethics and aesthetics are different projections of 'reality' providing apparently contradicting pictures which originate, however, from the same reality. Such an agreement was also reached with the Dalai Lama during a visit to CERN.

How such cultural values are to be quantified in dollars seems to be extremely difficult, even using kind of shadow prices.

### 5.2 Bringing nations together

The foundation of CERN was based on two initiatives: After the last World War European physicists recognised that Europe would only be competitive in science if the European forces are joined in an international laboratory. At the time politicians with a wide horizon were looking for a way to bring together European countries having been adversaries during the war, in peaceful work. A scientific laboratory was considered to offer an excellent possibility (A.Hermann et al. 1987) for such a purpose. The two initiatives were brought together at a meeting of UNESCO and CERN was founded under the umbrella of this organisation. As a result the constitution of CERN stipulates as its objectives not only the advancement of science but also bringing nations together, a task which today goes under the slogan 'science for peace' [6], CERN has fulfilled both these expectations in an extraordinary way. In the second case it has not only brought scientists from different nations and cultures together, but irradiates in many direct or indirect ways into politics. This can be exemplified by some anecdotes (Schopper 2009): during the hottest cold war CERN signed a contract with the Institute of High Energy Physics in the USSR which became a model for a similar contract between the USA and the USSR which in turn created sufficient trust for an agreement signed by the presidents Ford and Brezhnev. When the disarmament meeting between President Reagan and Gorbatchov was prepared in 1985 a physicist of the US delegation informed me that the official discussions got into a bottleneck and he asked whether CERN could provide the environment for a private discussion to unlock the deadlock and indeed an invitation to dinner at CERN, a neutral environment respecting both negotiating partners, unblocked the situation. For many decades the cooperation between CERN and the Joint Institute of Nuclear Research JINR at

[6] For a long time CERN was the only scientific organisation which had these objectives formally in its constitution. Recently the synchrotron radiation laboratory SESAME in Jordan is the second international laboratory with this dual function founded according to the CERN model. Indeed its constitution is more or less a copy of the CERN constitution

Dubna (USSR), a kind of CERN for the Warsaw Pact States, provided a unique link for scientific and personal contacts between West and East and it was the only possibility for contacts between the two Germanys. In a LEP experiment scientists from the Peoples Republic of China and Taiwan worked together for the first time in the same group which required approval from the highest governmental levels. In several cases the relations of CERN could be used to help dissidents. All these efforts and successes (extended from Europe to the whole world to a great extent thanks to LHC) were recognised in the most impressive way by the election of CERN as the only scientific organisation to advise the Secretary General of UN.

An evaluation of LHC, an integral part of CERN, is incomplete without taking into account the aspect ‘science for peace’. From a humanitarian point of view the resulting benefits may seem more important than most of the other economically oriented achievements.

### 5.3 A new style of practical international collaboration

Many large scientific projects are now realised in international or even global cooperation. Particle and nuclear physics have been leading this effort since many decades. At CERN a new style of global cooperation has been developed, partly voluntarily, partly pushed by necessity. At the beginning experiments at CERN were carried out by small groups comprising typically about a dozen scientists, most of the time originating in different countries. The equipment was mainly financed by CERN and outside users had only to cover their travel cost. Over the years when experiments became bigger and more complicated, some of the detector development was done at home laboratories and in special cases also the equipment was financed sometimes by the outside[7]. Under these conditions CERN could keep complete control over the construction and operation of the experiments.
This gradual transformation suffered an abrupt change when the LEP experiments where approved. The conditions for the LEP approval were such that practically no funds became available in the CERN budget for experiments (Schopper 2009). The total cost of all LEP experiments was about EUR 500 million, with about half of it from laboratories in CERN Member States and the rest from other, partly non-European countries. CERN could contribute only about 15% to this total cost. A large fraction of these contributions were provided in kind since most of the countries wanted that the components be manufactured at home, not only to save, but also to benefit from the resulting technological progress. As a result these so-called experiments became rather independent international organisations and a new way for their management had to be found.

The principles of these structures (‘experiments’) are:
A Steering Committee (now called Collaboration Board) is the decision taking body composed of institutional representatives of the various participating institutes. It elects for a fixed time a spokesperson. It decides on the scientific programme and on acceptance of new members.
Since the ideas and wishes of the scientists are sometimes too optimistic I insisted that for each LEP experiments a Finance Review Committee (for LHC now called Resource Review Committee) with representatives of funding agencies be installed for the financial supervision. They consider together with the Collaboration Board the overall financing of the project, the contributions in kind and the cash contributions to a common fund.
CERN keeps the right for the final decision to accept an experiment or refuse it. A Technical Coordinator is appointed by CERN to help to insure that the foreseen technical conditions are fulfilled and the time schedules met.

[7] Such an exception was the Big European Bubble Chamber BEBC, financed by France and Germany, but CERN could keep full control.

The ‘experiments’ have no legal identity, in particular no Chief Executive Officer.
Objectives are defined bottom-up by the collaboration. All partners are ‘equal’ (no leading country, no leading institution). Decisions are consensual, the main motivation is the success of the common project. There is no merciless competition, no shoot-out (European mentality versus US mentality). Cooperation and competition coexist! A new term was created ‘Co-opetition’ = co-operation + competition.
It is quite remarkable that truly world-wide projects with hundreds or even thousands of participants and cost of several hundred to billions of EUR could be realised successfully under such ‘liberal’ conditions respecting the time schedule and the budget forecasts.

What is the human and psychological basis for this success?
Cooperation in basic research may be considered as relatively easy: scientists are relatively rational, there is no secrecy, neither military nor industrial. However, also physicists are ambitious, but they realise that the success of the common project is also the basis for their personal career. Personal financial issues and status symbols are only secondary. Since decisions are taken collectively it is also appreciated that arbitrary or unfounded decisions can be largely avoided

Of course, it was not possible to introduce such a model in one go. It took several decades of experimenting and learning. It was first tried for the LEP experiments and further developed for the LHC experiments. It became the basis for the first USA participation in a project on non- US territory.

It is not trivial to make such projects a success as several other international projects have shown. The most drastic example was the failure of the Superconducting Supercollider SSC in the USA[8]. The comparison of the success of LHC with the failure of SSC has initiated discussions in the USA about the style of large international projects (M. Lucibella 2015). There Nigel Lockyer (director of Fermi Lab) is quoted as “CERN has done a great job of pulling together the world for their project, the LHC. It’s clearly been super-successful “and Lynn Orr (head of research at DOE) added “But it’s a success that is difficult to replicate”. Lockyer added that he and Ernest Moniz (US Secretary of Energy) have drawn the lesson that for a large project international partners need to be treated like real partners with real input into the project [9].

The CERN model, exemplified by LHC and its experiments, may provide a new paradigm for future large projects to be realised on a global level, perhaps even for industrial or other international projects. These achievements should certainly be considered as social benefits but how they could be evaluated in financial terms remains a mystery.

---

[8] This is not the place to discuss the different mistakes which led to its mischance, but one essential fault was that it never became clear whether it was a project to promote US superiority or a common global project. In addition the condition that all partners should be equal in the sense that they can influence the definition of the project independently of the size of their financial contribution was not respected. When the European and Japanese representatives were invited to negotiations at Washington we asked the question whether we still could discuss the final parameters of the project. The answer by J.Trievelpiece, the leader of the US delegation, was “the president has decided to build SSC, you join or leave it” (Schopper 2009). This was the end of the negotiations and in October 1993 the SSC was stopped by US Congress.

[9] He further added that “there are bad examples out there where that’s not been followed, where project management has been shown to be lacking.….You can look at the ITER project as an example”. The new director of ITER Bernard (see R. Gats, 2015) stated: The original cost tripled since the agreement was signed in 2007 and operation which was foreseen for 2016 may now perhaps start in 2023.

### 5.4 A special management style of CERN

In the previous chapter a new management style for individual large global projects was considered such as it evolved from the realisation of the LHC experiments. It is worthwhile to consider also the experience from the management of CERN as whole although the present CBA concerns only LHC. As has been explained in previous sections, LHC is so intimately integrated into CERN that the two can hardly be separated.

A special style for the overall CERN management had to be introduced in 1981 when the LEP project was approved by the CERN Council (Schopper 2009). Before 1981 the CERN budget was increased each time a new big project had been approved. However, when in 1981 the two CERN laboratories (CERN I and CERN II existed with two Director-Generals rather independently since the construction of the SPS)[10] were united, a total budget was approved which was lower than the sum of the previous two budgets (Fig.2).

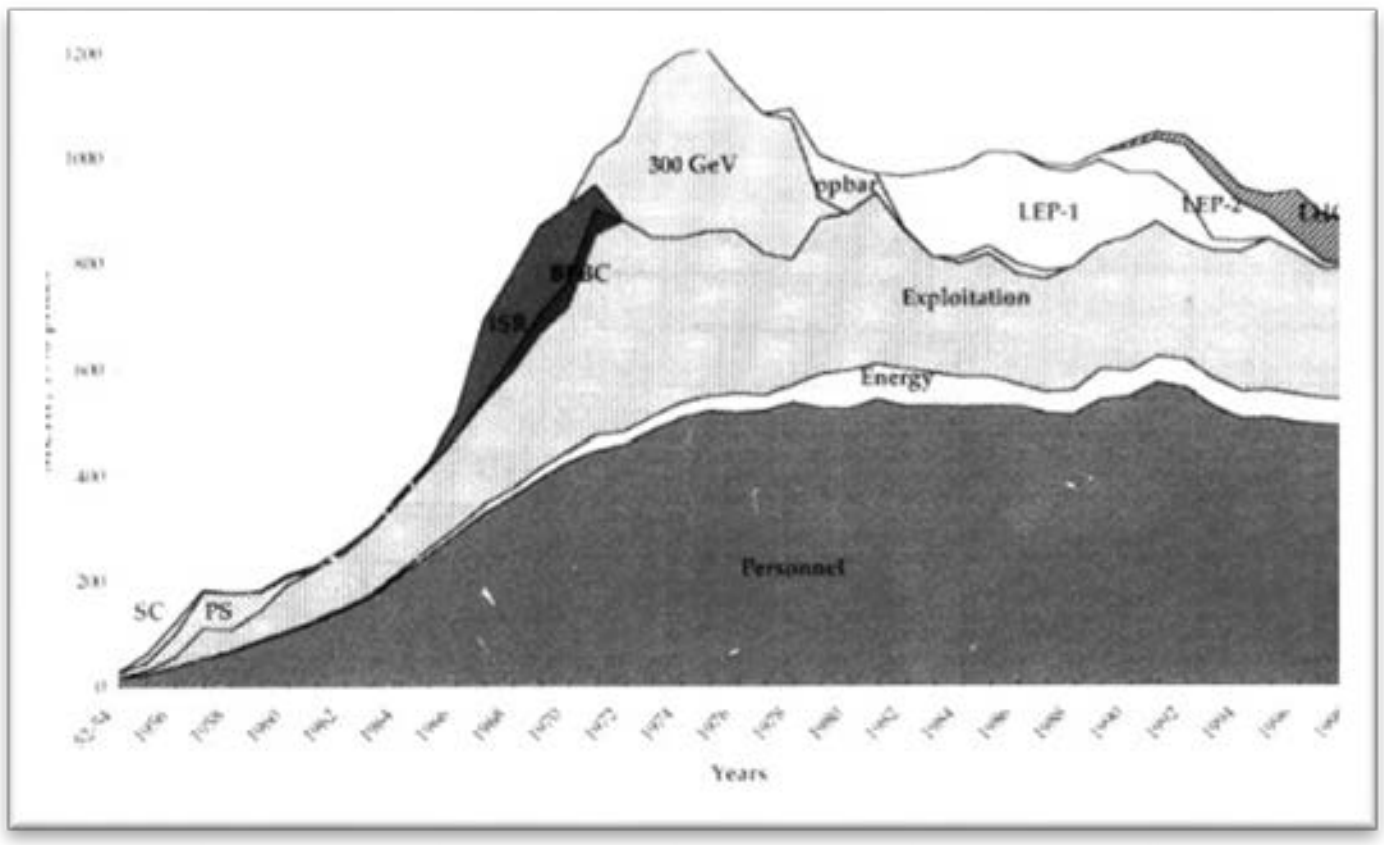

**Figure 2. Total CERN Budget (CERN I + CERN II) 'Constant' since 1981**

The approval of LEP was linked to three additional conditions:

- all Member States had to participate in the LEP project which was considered to be part of the 'basic programme' of CERN (with the consequence that no Member State must vote against the project), a condition which was automatically extended to LHC,
- a definite total cost was mentioned for LEP at its approval, but it was understood that the project had to be constructed within the total budget of CERN (Fig.3), no matter what kind of unexpected events would occur. We devised the slogan 'time is our contingency',
- Council required a reduction of the CERN personnel by an early retirement programme without approving the necessary compensation for the losses of the Pension fund (Fig.5).[11]

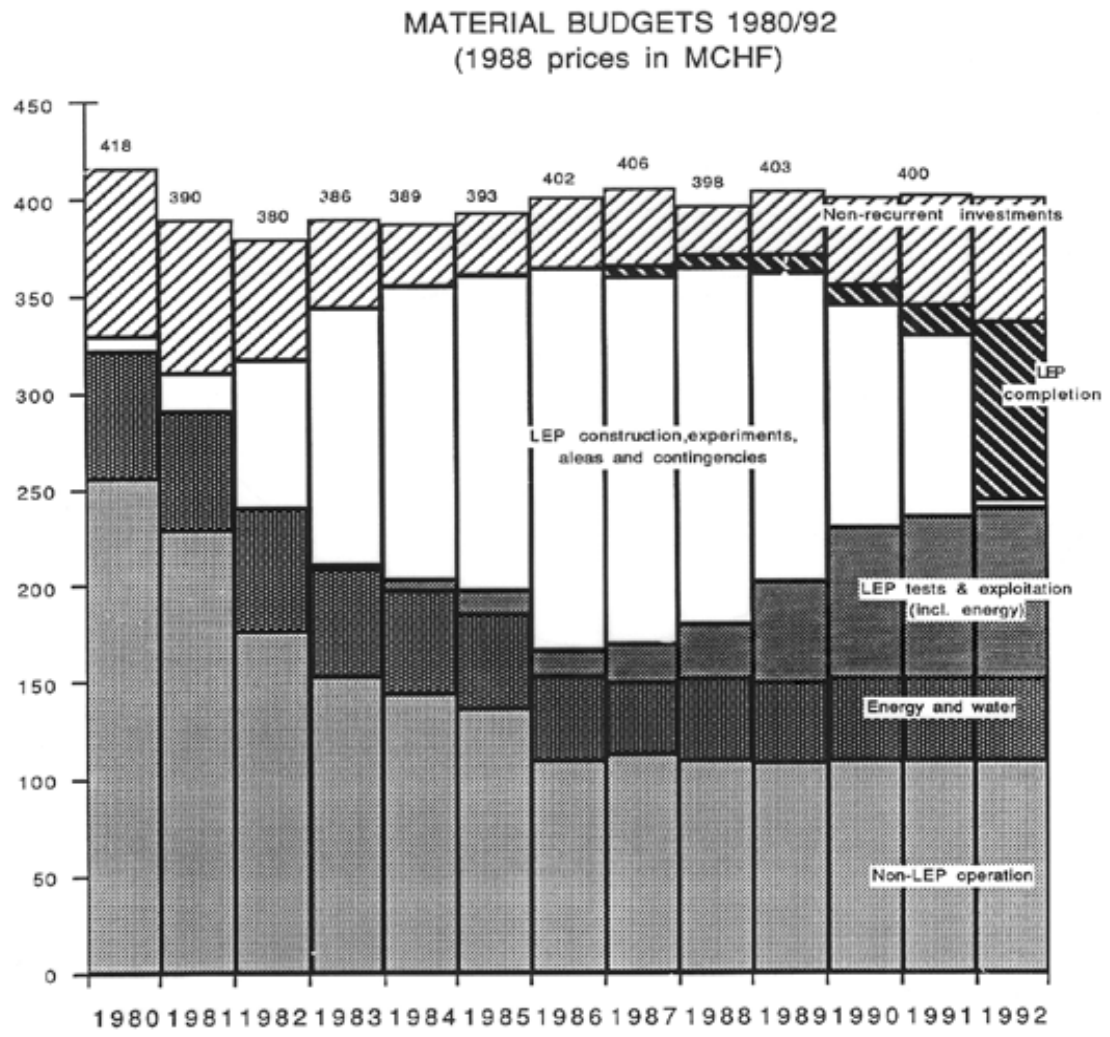

**Figure 3. The CERN Budget for material resources during LEP construction**

---

[10] Originally it was planned to build SPS in a different laboratory CERN II on a different site. Fortunately this idea was given up, but a separate organisation was maintained. The 'unification' of CERN with a constant budget meant in reality a considerable reduction of resources. Many still unique programmes (e.g.ISR) had to be closed down, a large part of the infrastructure had to be restructured and about 1/3 of CERN staff had to be reemployed for LEP.

[11] This problem is still not fully settled by today but does not directly concern the C/BA of LHC.

These conditions were more or less still valid after 30 years during the LHC construction, with both their advantages and disadvantages. The advantage is that within the total budget the Council leaves a remarkable freedom to the CERN management how to execute the programme. There is almost no micro- management by Council and its committees. In particular a transfer of funds from investment to operation is possible, a freedom which is unthinkable in some countries where the parliaments decide independently on investment and operation. Only with such a flexibility it became possible to realise LEP and LHC within a fixed total budget[12] . Thanks to this practice of dealing with the total budget, CERN was also spared the fluctuations in supporting scientific research as it occurred in some national budgets. This provided a certain stability and predictability of CERN

The big disadvantage was that the total budget was fixed at a level considerably lower than asked for by the Director-General and LEP could be built only by terminating some of the most interesting and unique projects of CERN. The LEP budget had to be 'carved out' of the existing budget for material ( Fig.3). Even then it was not possible to accommodate the profile of yearly expenditures for LEP which display a kind of bell-shape (Fig. 4), within a constant budget. For the peak payments in about the middle of the project loans had to be made. The Council refused bank loans and the only solution was a loan from the CERN pension fund guaranteeing a relatively high interest rate. Since this problem of accommodating the yearly expenses of a project within a constant total budget still existed during the LHC construction the same difficulty appeared again. This time the Council agreed that CERN took a bank loan which still has to be paid back.

Finally as has been mentioned any unforeseen expenses have to be covered within the constant budget. This was true for the extra cost of the water incident of the LEP tunnel (see chapt. 2) and also the consequences of a LHC accident had to be settled within the constant budget.
For an institution which is operated under such conditions it might seems more reasonable to evaluate the whole organisation instead of a particular project.

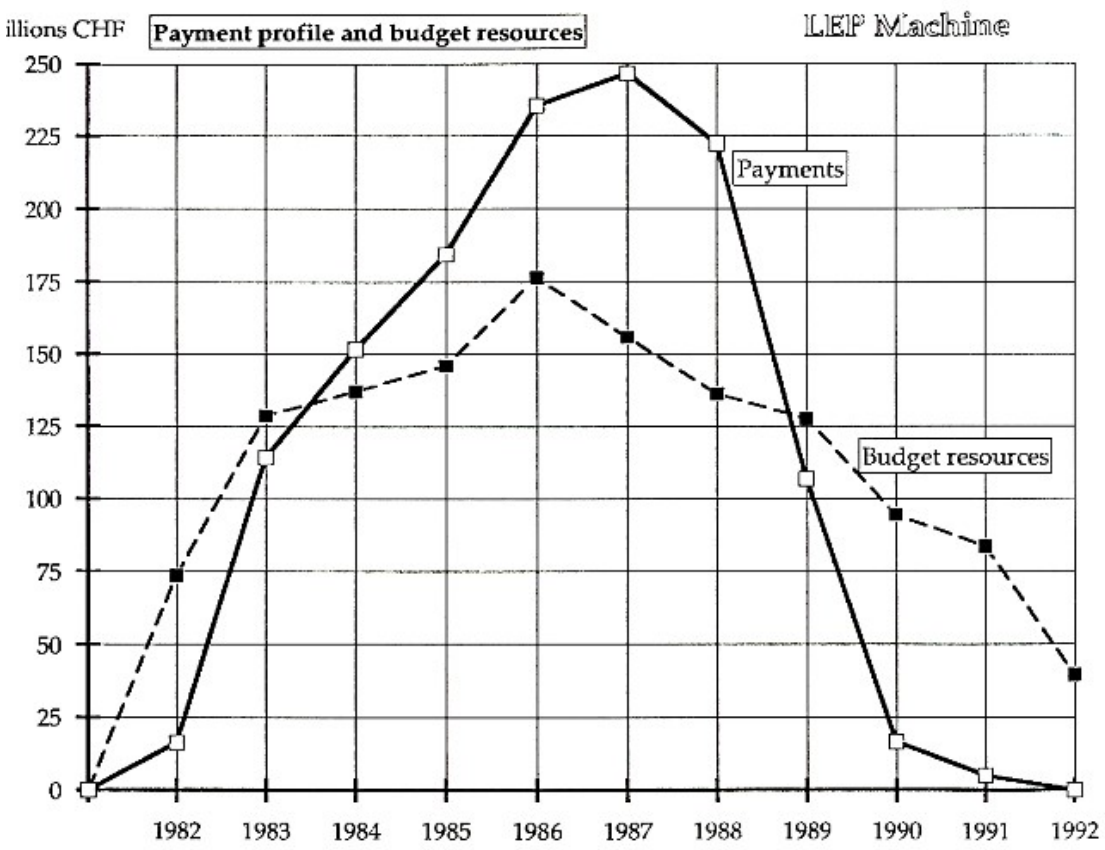

**Figure 4. Payments profile for LEP machine**

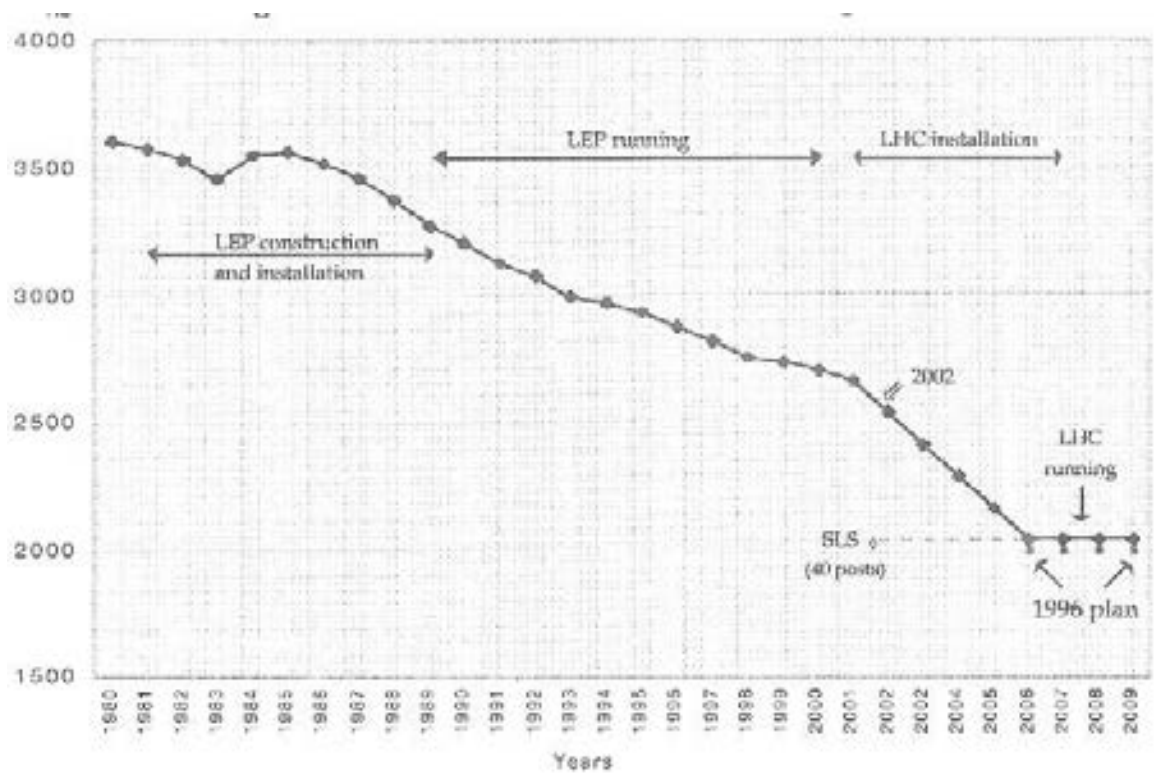

**Figure 5 Reduction of CERN personal with early retirement scheme machine**

---

[12] There was a partial compensation for inflation and minor corrections after the joining of new Member States.

## 6. Conclusions

Scientific progress and technical innovations are increasingly considered as necessary elements for economic growth. The growing cost of large research infrastructures, also for basic research, make it more difficult for politicians and funding agencies to take decisions since the benefits for society can hardly be known *ex-ante*. Hence to avoid possible criticism decision takers try to justify their decisions as much as possible by 'objective' criteria. Growing cost on one side and more serious constraints for public expenditure require a better public accountability and a more rigorous assessment of projects of basic research (Martin 1996).

Cost Benefit Analysis can be one element in assessing scientific infrastructures. This methodology has been well developed for industrial projects (e.g. energy, environment, transport) and now Massimo Florio and his collaborators have made a big step forward in applying this methodology to basic scientific infrastructure (Florio et al, 2015). They have chosen the LHC at CERN as a study case to test the method and they find a considerable social benefit for this project. In spite of an immense effort to assess the cost and benefits of LHC as objectively as possible they are conscious of the present limitations of the method and suggest further research. In this paper some arguments are given to corroborate the difficulties in applying C/BA to basic science projects and the hope is that further studies might be incited.

One of the main weaknesses is that a precise definition of some of the used variables (both cost and benefits) is extremely difficult and causes many ambiguities. One drastic example is the evaluation of the scientific benefits of LHC which by the C/BA is given as less than 2% of the total benefits, a drastic understatement. Such uncertainties are partly taken into account by treating all the variables as stochastic and calculating probability distributions. But the result is that the final errors of some of the individual costs or benefits are larger than the final *Net Present Value* which, being positive, provides the final test for a project. This implies that at the present stage of the methodology the *NPV* alone should not be used to find out whether 'a project passes the CBA test'. However, the specific information provided by C/BA on individual benefits or costs can be very useful to evaluate independently different aspects of a project (e.g. technology transfer, education).

In addition to these difficulties which are inherent to the present C/BA method as applied to basic science infrastructures there are some aspects special for LHC and CERN. One particular aspect is the intimate integration of LHC in existing or previous projects which makes the separation of costs extremely difficult (e.g. the cost of general administration. the value of existing experience or even the tunnel cost).

There exist other elements crucial for CERN for which no way of integration into a CB analysis can be seen at present.

CERN was founded with the objectives to promote science but also to foster the understanding between nations. An evaluation of the social value of a project like LHC cannot be complete without considering this aspect of 'science for peace'. The election of CERN as an advisory organisation to UN is only one recognition of these achievements.
Another element is the establishment of a new style of global international cooperation, both by the 'LHC experiments' (in reality large independent projects), but also how CERN as a whole is managed.
In a world where the complexity of global actions increases continuously, the exploration of new ways of managing new challenges has a considerable social value independent of CERN's scientific success. Hence it is not surprising that the CERN model has been discussed several times at World Economic Forum at Davos.

In spite of the considerable progress which has been achieved for the C/BA method applied to basic research infrastructures, a warning is necessary in view of the large inherent uncertainties. If such an analysis is used by decision makers who do not have the time to study its details and benefits for partial results, but would be tempted to look just at the final outcome (the *Net Present Value*), wrong conclusions could be drawn. CBA can be a useful element for decisions taking, but should not be considered as the only or even the main instrument.

Acknowledgement: Prof. M.Florio and his collaborators I should like to thank for many interesting and instructive discussions.